\documentstyle[twocolumn,aps]{revtex}
\input epsf.tex
\begin{document}
\title{Orbital Kondo Effect in Ce$_x$La$_{1-x}$B$_6$: Scaling Analysis}
\author{H. Kusunose\cite{kusu} and Y. Kuramoto}
\address{Department of physics, Tohoku University, Sendai, 980-8578, Japan}
\date{\today}
\maketitle
\begin{abstract}
Peculiarity of the Kondo effect in Ce$_x$La$_{1-x}$B$_6$ is investigated on the basis of the scaling equations up to third order.
For the case where the $f^1$-$f^2$ charge fluctuation enters in addition to the $f^1$-$f^0$ one, the effective exchange interaction becomes anisotropic with respect to the orbital pseudospins which represent the two different orbitals in the $\Gamma_8$ ground state.
Because of different characteristic energies for electric and magnetic tensors, scaling with the single Kondo temperature does not apply to physical quantities
such as the resistivity and magnetic susceptibility.
Possibility of a bizzare phase is pointed out where the RKKY interaction leads to the spin ordering without orbital ordering.
This phase serves as a candidate of the phase IV which is observed to be isotropic magnetically.
\end{abstract}

\narrowtext
%
%
\section{Introduction}
Orbital dynamics in a number of heavy-fermion systems has recently attracted great interest \cite{cox,koga,uimin,shiina,kuramoto,fukushima}.
In the presence of orbital degeneracy, orbital (electric) tensors in addition to spin ones have an opportunity to be active, and shows rich phenomena resulting from entanglement with spin (magnetic) degrees of freedom.
The orbital degrees of freedom couple with lattice ones, and often lead to the Jahn-Teller effect.
In other cases, a quenching mechanism for these degrees of freedom such as the orbital Kondo effect prevents static distortion of the lattice.
Hence study of the coupled spin and orbital fluctuations should be important for
understanding heavy-fermion systems as a whole.

Under a high symmetry such as the cubic one, there are chances for the orbital degeneracy to remain.
A typical example is the cubic compound CeB$_6$: its crystal-field ground state is the $\Gamma_8$ quartet which consists of two degenerate Kramers doublets.
The excited doublet $\Gamma_7$ is well separated by about 540K \cite{zirngiebl} and plays little role in low-energy physics.
CeB$_6$ exhibits curious phase diagram at low temperature in magnetic field \cite{effantin,kawakami,takigawa}.
The phase boundary between the paramagnetic phase, called I, and the antiferro-quadrupolar phase called II shows unusual dependence on magnetic field: transition temperature increases as magnetic field increases.
The magnetic field dependence of the phase II has been ascribed either to the intersite interactions between higher-order multipoles \cite{ohkawa,shiina}, or the quadrupolar fluctuations \cite{uimin,fukushima}.

In the course of systematic dilution study \cite{nakamura,tayama,hiroi}, a strange phase, called IV, is observed recently in Ce$_x$La$_{1-x}$B$_6$.
The magnetic susceptibility shows a cusp on entering the phase IV from the paramagnetic phase I with decreasing temperature.
This suggests that the N\'eel state is present here.
In contrast to the phase III which has both antiferromagnetic and quadrupolar orders, the phase IV has a very small magnetic anisotropy in the susceptibility\cite{tayama} and almost no magneto resistance\cite{hiroi}.
In realizing this phase the interplay between intersite correlation and on-site Kondo effect seems to be essential.

In a previous paper \cite{kuramoto2}, one of the present authors noted the importance of the orbital Kondo effect in understanding the phase IV; the orbital Kondo effect can be active even in the presence of a spin ordering.
Then a new method was proposed to perform perturbative renormalization to arbitrary higher orders.
In order to quantify the idea, however, one must estimate actual magnitude of relevant interactions under the realistic point-group symmetry.
Up to the present, investigation of the Kondo-type interaction in the presence of crystalline-electric-field (CEF) effects has been made by various methods \cite{cox,koga,hirst,kim}.
It has turned out essential to take into account the splitting of the localized states in accordance with the point-group symmetry \cite{cox}.
As a result, the effective exchange interaction needs many parameters for characterization.
In the conventional field theoretical scaling procedure, it is tedious to deal with such large number of parameters.

In this paper we generalize the scaling method of Ref. \cite{kuramoto2} so as to be applicable to arbitrary point-group symmetry.
The scaling equations for symmetry adapted coupling constants are written down generally in terms of structure constants of the relevant Lie algebra.
As a specific case, the third-order scaling for the cubic symmetry is performed explicitly with minimum amount of intermediate steps.
We show that the pseudospin representing the orbital moment has an anisotropic exchange interaction as a result of scaling.
On the other hand the magnetic pseudospin remains isotropic in the case where the $f^1$-$f^0$ fluctuation dominates the $f^1$-$f^2$ one.
Since the third-order scaling does not work for a strong-coupling fixed point, the local Fermi liquid cannot be identified by the present approach as it stands.
However, with available knowledge from various sources, we can almost certainly classify all the fixed points of the model.
It turns out that there is an unexpected symmetry in the exchange interaction which interchanges both the spin and orbital indices simultaneously.
We show that this hidden symmetry is specific to the SU(2)$\times$SU(2) symmetry which is relevant to the $\Gamma_8$ CEF state.

This paper is organized as follows:
In the following section, we describe the method to derive the effective exchange interaction starting form the Anderson model.
This section is mainly adaptation of earlier treatment \cite{hirst} to respect the point group symmetry from the beginning.
The details of explicit derivation are given in Appendix A.
In section III we derive the third-order scaling equation in the most general case of the point group.
In section IV we apply the general result to the specific case of the Ce$_x$La$_{1-x}$B$_6$ system and  give detailed analysis of the scaling equations.
The final section summarizes the present paper with implication to identifying the mysterious phase IV.

%
%
\section{Exchange Interaction in Irreducible Representation}
The atomic structure of a single magnetic ion with $f$ electrons is treated by the CEF theory together with the Russell-Saunders ($LS$) coupling scheme.
In this scheme the local state is specified uniquely by the following quantum numbers:
the number of $f$ electrons $f^n$, the orbital, the spin and the total angular moments $L,S,J$, and the irreducible representation $\Gamma$ (abbreviated as irrep) of the double point-group with time-reversal operation and its component $\gamma$ together with the branching multiplicity label $Q$, which is required for the case where the irrep $\Gamma$ occurs more than once (e.g. $D_{5/2}=\Gamma_6\oplus 2\Gamma_7$ for tetragonal symmetry).
We often abbreviate the local states $|f^nLSJQ\Gamma\gamma\rangle$ as $|\phi_n\gamma\rangle$ for notational simplicity.
In the same manner the one-particle state of $f$ electron is described as $|\xi\lambda\rangle$ where $\xi$ is the abbreviation of $(\ell=3,s=1/2,jq\Lambda)$ with $q$ being the multiplicity label, and $\Lambda$ specifying the irrep of the one-particle state.
The corresponding creation operator is written as $f^\dagger_{\xi\lambda}$.
It is useful to express a Bloch state conduction operator $c^\dagger_{{\bf k}\sigma}$ using $c^\dagger_{k\xi\lambda}$ with the symmetry adapted basis around the impurity

The Anderson Hamiltonian then takes the form:
\begin{eqnarray}
&& H=H_k + H_f + H_{\rm hyb},\label{eqand}\\
&& \mbox{\hspace{.5cm}} H_k =
\sum_{k\xi\lambda}\epsilon_{k\xi}c^\dagger_{k\xi\lambda}c_{k\xi\lambda},\\
&& \mbox{\hspace{.5cm}} H_f =
\sum_{\phi_n\gamma}E_f(\phi_n)|\phi_n\gamma\rangle\langle\phi_n\gamma|,\\
&& \mbox{\hspace{.5cm}} H_{\rm hyb} =
\sum_{k\xi}\sum_{\lambda}\left[V_{k\xi}c^\dagger_{k\xi\lambda}f_{\xi\lambda} + {\rm
h.c.}\right],
\end{eqnarray}
where we restrict the hybridization between the local and the conduction electrons to the same set of symmetry (not only $\Lambda_c=\Lambda_f$ but also $\xi_c=\xi_f$).
We regard the label $\xi$ as the channel index of independent scattering processes and $\lambda$ as the internal degrees of freedom which is responsible for the Kondo effect.

\begin{figure}
\begin{center}
\epsfxsize=8cm \epsfbox{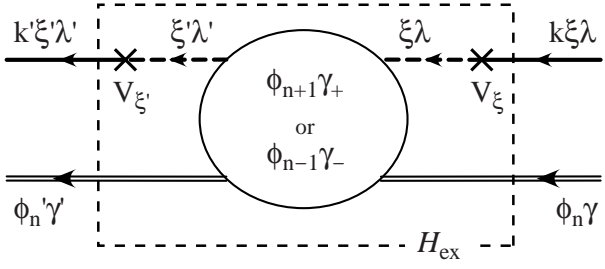}
\end{center}
\caption{The exchange process via the excited configurations $\phi_{n\pm1}\gamma_{\pm}$.
The solid (dashed) line denotes the conduction (one-particle $f$) electron.
The double solid line represents the localized state in the projected space.}
\label{fig1}
\end{figure}

If the most stable configuration $f^n$ is well separated from $f^{n\pm1}$ ones, one can restrict the model space \cite{lindgren} to the multiplet $(f^nLSJ)$ by integrating out the virtual charge fluctuations to the $f^{n\pm1}$ configurations.
The effective Hamiltonian in the model space is written as
\begin{eqnarray}
&& H_{\rm eff} = H_k + {\cal P}H_f{\cal P} + H_{\rm ex} + {\cal O}(V^4),\\
&& H_{\rm ex} = \sum_{a,b\in {\cal M}}|a\rangle\langle a|V(E_b-H_f)^{-1}{\cal
Q}V|b\rangle\langle b|,
\end{eqnarray}
where ${\cal P}$ denotes the projection operator to the model space ${\cal M}$ and ${\cal Q}=1-{\cal P}$.
We have neglected $H_k$ in the resolvent of intermediate state assuming that the relevant conduction-electron states have energies smaller than those for excited states of $H_f$.
The exchange process via the excited configurations is shown schematically in Fig.\ref{fig1}.
Then the exchange interaction generally takes the form
\begin{equation}
H_{\rm ex}=\sum_{i\alpha} J_{i\alpha}\;c^\dagger {\sf x}^i c\cdot|\phi '\rangle {\sf X}^{\alpha}\langle\phi|,
\label{eq1}
\end{equation}
where meaning of the indices $i$ and $\alpha$ are to be specified later.
The matrices ${\sf x}^i$ and ${\sf X}^{\alpha}$ describe the transition processes for conduction electrons and the localized states, respectively.
Summation over quantum numbers of each state is implied by the matrix multiplication.
The numbers of independent matrices are $14^2-1$ for the conduction electron and $d_J^2-1$ for the multiplet $(f^nLSJ)$ with the degeneracy $d_J$.
The unit matrices which give the potential scattering do not enter the exchange Hamiltonian.

The invariance under the point-group operation is implicit in eq.(\ref{eq1}).
To make it explicit \cite{hirst}, we introduce the irreducible tensor operators \cite{butler} defined as follows:
\begin{eqnarray}
&&
x^{(r)}_{\Delta^*\delta^*}(\xi'\xi)=\sum_{k'\lambda'k\lambda}c^\dagger_{k'\xi'\lambda'}
c_{k\xi\lambda}\left(\begin{array}{c} \Lambda' \\ \lambda' \end{array}\right)
\left(\begin{array}{ccc} \Lambda'^* & \Delta^* & \Lambda \\ \lambda'^* & \delta^* & \lambda
\end{array}\right)_r,\nonumber\\
&&
X^{(t)}_{\Delta\delta}(\phi_n'\phi_n)=\sum_{\gamma'\gamma}|\phi_n'\gamma'\rangle
\langle\phi_n\gamma|
\left(\begin{array}{c} \Gamma' \\ \gamma' \end{array}\right)
\left(\begin{array}{ccc} \Gamma'^* & \Delta & \Gamma \\ \gamma'^* & \delta & \gamma
\end{array}\right)_t,\label{eqten}
\end{eqnarray}
where the one-column and the three-column brackets indicate $2jm$ and $3jm$ symbols with the multiplicity $r$ or $t$\cite{butler}.
These are natural extensions of the $1j$ and $3j$ symbols of Wigner.
It is clear that the tensor operator $x^{(r)}_{\Delta^*\delta^*}(\xi'\xi)$ transforms like the ket $|r\Delta^*\delta^*\rangle$, or equivalently like the bra $\langle r\Delta\delta|$, under the point-group operations.
The irreps of the localized-state tensors are determined so as to have the finite $3jm$ symbols.
Namely, decompositions of the direct product $\Gamma'^*\otimes\Gamma$ contains $\Delta$.
Similar decomposition also determines the irreps of the conduction-electron tensors.
We write the exchange interaction $J_{i\alpha}$ as another matrix $g$ in the invariant form.
Namely eq.(\ref{eq1}) is equivalently written as
\begin{eqnarray}
&&H_{\rm ex} =
\sum_{\xi'\xi\phi_n'\phi_n}\sum_{\Delta rt}
g^{(rt)}_{\Delta}(\xi'\xi;\phi_n'\phi_n)
\nonumber\\&&\mbox{\hspace{2cm}}\times
\sum_\delta
\left(\begin{array}{c} \Delta \\ \delta \end{array}\right)
x^{(r)}_{\Delta^*\delta^*}(\xi'\xi)X^{(t)}_{\Delta\delta}(\phi_n'\phi_n).
\label{eqexe}
\end{eqnarray}
Here the $2jm$ symbol  is inserted in the exchange interaction.
With the rotational symmetry, eq.(\ref{eqexe}) reduces to the known result involving the spherical tensors \cite{hirst,koga}.

The magnitude of the coupling constants, which will be obtained explicitly in Appendix A, are given by
\begin{eqnarray}
&&g^{(rt)}_{\Delta}(\xi'\xi;\phi_n'\phi_n)
\nonumber\\&&
=\sum_{\phi_{n+1}}
A^{(rt)}_{\Delta}(\phi_{n+1})\langle\phi_n'||f_{\xi'}||\phi_{n+1}\rangle\langle\phi_{n+1}||f^
\dagger_\xi||\phi_n\rangle
\nonumber\\&&
+\sum_{\phi_{n-1}}
A^{(rt)}_{\Delta}(\phi_{n-1})\langle\phi_n'||f^\dagger_\xi||\phi_{n-1}\rangle\langle\phi_{n-1
}||f_{\xi'}||
\phi_n\rangle,\label{eqsel}
\end{eqnarray}
where $A^{(rt)}_{\Delta}(\phi_{n\pm1})$ is of the order of $|V|^2/[E_f(\phi_{n\pm1})-E_f(\phi_n)]$.
The first summation is taken for the excited configuration $f^{n+1}$ such that both the direct products $\Gamma'\otimes\Lambda'\otimes\Gamma_+$ and $\Gamma_+\otimes\Lambda\otimes\Gamma$ contain the identity representation.
Similar selection rule is available to the $f^{n-1}$ configuration \cite{cox2}.

Hereafter we restrict our discussion for simplicity to the case where the irrep $\Delta\delta$ is real and the corresponding $2jm$ symbol can be set to unity.
Moreover, both the initial and the final states belong to the ground-state CEF multiplet $\phi_g$.
Then the localized-state tensors become Hermitian and all have definite signs with respect to the time-reversal operation.
The conduction-electron tensor does not have the definite sign in general, since the time-reversal operation interchanges $\xi'$ with $\xi$.
However, if one considers the matrix of the tensors in the combined space $\xi'\oplus\xi$ which are defined as
\begin{equation}
\hat{x}^{(r)}_{\Delta\delta}(\xi'\xi)=\left\{
\begin{array}{ll}
x^{(r)}_{\Delta\delta}(\xi'\xi) & (\xi'=\xi) \\
x^{(r)}_{\Delta\delta}(\xi'\xi)+\left[x^{(r)}_{\Delta\delta}(\xi'\xi)\right]^\dagger & ({\rm otherwise}),
\label{eqsym}
\end{array}\right.
\end{equation}
then the matrix of the tensors have definite signs under  the time reversal \cite{hirst}.
In the above discussion we use the fact that the coupling constants are real and symmetric against interchange of $\xi'$ and $\xi$ which is ensured by the hermiticity and the time-reversal symmetry of the exchange interaction.
In the treatment above the same coupling constant is automatically imposed for the different channels.
This property was essential to derive the two-channel Kondo Hamiltonian for tetragonal and hexagonal symmetries \cite{cox2}.
We rewrite the exchange interaction in the restricted case as
\begin{equation}
H_{\rm ex} = \sum_{\langle\xi'\xi\rangle}\sum_{\Delta rt}
g^{(rt)}_{\Delta}(\xi'\xi;\phi_g)\sum_\delta {\hat
x}^{(r)}_{\Delta\delta}(\xi'\xi)X^{(t)}_{\Delta\delta}(\phi_g),
\label{eq2}
\end{equation}
where the summation of channels is taken for combination of a pair of $\xi'$ and $\xi$ because of the symmetrized expression introduced by eq.(\ref{eqsym}).

%
%
\section{Scaling Equations for Generalized Exchange Interaction}
We derive the scaling equations up to third order for the exchange interaction in terms of the irreducible tensors.
The index $i$ in eq.(\ref{eq1}) is abbreviation of $(r\Delta\delta,\xi'\xi)$ and the index $\alpha$ of $(t\Delta\delta,\phi_g)$.
In dealing with the matrices representing the irreducible tensor operators, we note that the matrix ${\sf x}^i$ actually depends only on a subset of quantum numbers $(\Delta,\Lambda',\Lambda)$, and the matrix ${\sf X}^\alpha$ on $(\Delta,\Gamma_g)$ where $\Gamma_g$ is the irrep of $\phi_g$.
We write the dimension $d(\Delta,\Lambda',\Lambda)$ of ${\sf x}^i$ simply as $d_i$ and the dimension $D(\Delta,\Gamma_g)$ of ${\sf X}^\alpha$ as $D_\alpha$.
To treat all matrices on an equal footing, each matrix should be embedded in a space of $14\times14$ matrix.
The exchange $J_{i\alpha}$ in eq.(\ref{eq1}) is defined as $J_{i\alpha}=g^{(rt)}_\Delta(\xi'\xi;\phi_g)$ for $(\Delta_c\delta_c)=(\Delta_f\delta_f)$ and zero otherwise.
Here $J_{i\alpha}$  does not depend on the component index $\delta$.

The matrices satisfy the orthogonality relation:
\begin{eqnarray}
&& {\rm Tr}({\sf x}^i{\sf x}^j)=d_i\delta_{ij},\nonumber\\
&& {\rm Tr}({\sf X}^\alpha {\sf X}^\beta)=D_\alpha\delta_{\alpha\beta}.
\label{eqno}
\end{eqnarray}
The commutation rule is given by
\begin{eqnarray}
&& [{\sf x}^i,{\sf x}^j]=i\sum_k \frac{1}{d_k}  f_{ijk}{\sf x}^k,\\
&& [{\sf X^\alpha},{\sf X^\beta}]=i\sum_{\gamma}\frac{1}{D_\gamma} F_{\alpha\beta\gamma}{\sf X^\gamma}.
\end{eqnarray}
Equivalently the structure constant is given explicitly by
\begin{eqnarray}
&& f_{ijk}=-i{\rm Tr}([{\sf x}^i,{\sf x}^j]{\sf x}^k),\label{eqf1}\\
&& F_{\alpha\beta\gamma}=-i{\rm Tr}([{\sf X^\alpha},{\sf X^\beta}]{\sf X}^\gamma).
\label{eqf}
\end{eqnarray}
It is obvious that the structure constants $f_{ijk}$ and $F_{\alpha\beta\gamma}$ are completely antisymmetric against interchange of a pair of indices.

According to the renormalization formalism based on the open-shell Rayleigh-Schr\"odinger perturbation theory \cite{kuramoto2}, a change of the band cut-off $E_c$ induces the following expansion of the effective interaction matrix:
\begin{equation}
{\sf h}_{\rm int} = {\sf h}_{\rm ex} + \delta {\sf h}_{\rm int} ^{(2)}+ \delta {\sf h}_{\rm int}^{(3)}+\cdots,
\end{equation}
where the superscript indicates the order of the bare coupling constant, and the lowest-order matrix ${\sf h}_{\rm ex}$ is given by
\begin{equation}
{\sf h}_{\rm ex}=\sum_{i\alpha}J_{i\alpha}{\sf x}^i{\sf X}^\alpha.
\end{equation}

\begin{figure}
\begin{center}
\epsfxsize=8cm \epsfbox{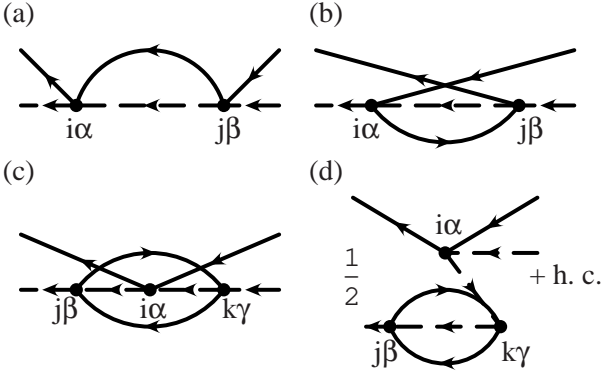}
\end{center}
\caption{Scattering processes (a), (b) in second order and (c), (d) in third order.
The solid line shows a conduction-electron state, while the dashed line the local electron.
The backward propagation of the dashed line in (d) is characteristic of the folded diagram.
Each index denotes the corresponding coupling constant.
The intermediate conduction-electron states is required to have energies near the cut-off $E_c$.}
\label{fig2}
\end{figure}

The second-order contribution shown in Fig.\ref{fig2}(a) and \ref{fig2}(b) to the effective interaction is given by
\begin{eqnarray}
&&\delta{\sf h}_{\rm int} ^{(2)}/\delta(\ln E_c)=\sum_{ij}\sum_{\alpha\beta}J_{i\alpha}J_{j\beta}[{\sf x}^i,{\sf x}^j]({\sf X}^\alpha {\sf X}^\beta)
\nonumber\\&&\mbox{\hspace{2cm}}
= \frac{1}{2}\sum_{ij}\sum_{\alpha\beta}J_{i\alpha}J_{j\beta}[{\sf x}^i,{\sf x}^j][{\sf X}^\alpha,{\sf X}^\beta],
\end{eqnarray}
where we have interchanged the dummy indices $(i,\alpha)\leftrightarrow(j,\beta)$ in deriving the second equality.

Figure \ref{fig2}(c) shows a diagram for the third-order contribution to the effective interaction.   One should also take into account the ``folded diagram''\cite{kuramoto2} shown in Fig.\ref{fig2}(d).
Taking average of the original and Hermite-conjugate counterpart of the folded diagrams, we obtain
\begin{eqnarray}
&&\delta{\sf h}_{\rm int}^{(3)}/\delta(\ln E_c)=\frac{1}{2}\sum_{ijk}\sum_{\alpha\beta\gamma}
J_{i\alpha}J_{j\beta}J_{k\gamma}{\sf x}^i{\rm Tr}({\sf x}^j{\sf x}^k)
\nonumber\\&&\mbox{\hspace{3cm}}\times
\{{\sf X}^\beta[{\sf X}^\gamma,{\sf X}^\alpha]-[{\sf X}^\beta,{\sf X}^\alpha]{\sf X}^\gamma\}
\nonumber\\&&\mbox{\hspace{0.5cm}}
=\frac{1}{2}\sum_{ijk}\sum_{\alpha\beta\gamma}J_{i\alpha}J_{j\beta}J_{k\gamma}{\sf x}^i{\rm Tr}({\sf x}^j{\sf x}^k)[{\sf X}^\beta,[{\sf X}^\gamma,{\sf X}^\alpha]].
\end{eqnarray}
Here we have also interchanged the dummy indices $(j,\beta)\leftrightarrow(k,\gamma)$ in deriving the second equality.

By computing the commutators we find that the ${\sf h}_{\rm int}$ has the same matrix structure as the lowest order part ${\sf h}_{\rm ex}$.
Thus, we obtain the scaling equations in terms of the structure constants as
\begin{eqnarray}
&&\frac{\partial}{\partial\ell}J_{a\epsilon}=\beta^{(2)}_{a\epsilon}+\beta^{(3)}_{a\epsilon},
\nonumber\\&&\mbox{\hspace{1cm}}
\beta^{(2)}_{a\epsilon}=-\frac{1}{2d_aD_\epsilon}\sum_{ij}\sum_{\alpha\beta}J_{i\alpha}
J_{j\beta}f_{ija}F_{\alpha\beta\epsilon},
\nonumber\\&&\mbox{\hspace{1cm}}
\beta^{(3)}_{a\epsilon}=\frac{1}{2D_\epsilon}\sum_{\alpha\beta\gamma}J_{a\alpha}\sum_{j}d_jJ_{j\beta}
J_{j\gamma}\sum_{\delta}\frac{1}{D_\delta}F_{\alpha\gamma\delta}F_{\epsilon\beta\delta},
\nonumber\\
\label{eqse}
\end{eqnarray}
with $\ell=\ln E_c$.
This is the most general form of the third-order scaling which is valid for any point-group symmetry.
We emphasize that non-commuting property of tensor operators is concisely taken into account in terms of structure constants of the underlying Lie algebra.

%
%
\section{Application to Cubic Symmetry}
In this section the scaling analysis developed  in the previous sections is applied to the case of dilute system Ce$_x$La$_{1-x}$B$_6$ which exhibits the remarkable entanglement of magnetic and electric tensors in static and dynamic properties.
We first derive the exchange interaction with use of pseudospins and then discuss nature of scaling.

\subsection{Exchange Interaction}
The magnetic ion Ce$^{3+}$ ($f^1$) lies in the cubic-symmetry ($O_h$) CEF.
The degeneracy of the ground multiplet $^2F_{5/2}$ is lifted to the excited doublet $\Gamma_7$ and the ground quartet $\Gamma_8$.
The partial waves of conduction electron are also classified by the cubic symmetry:
\begin{eqnarray}
&&D_{5/2}=\Gamma_7\oplus\Gamma_8,\\
&&D_{7/2}=\Gamma_6\oplus\Gamma_7\oplus\Gamma_8.
\end{eqnarray}
We define the basis sets for both conduction $(j)$ and localized $(J)$ states as follows:
\begin{flushleft}
(i) $j,\;J=5/2$
\end{flushleft}
\begin{eqnarray}
&&|\Gamma_7:(\uparrow,\downarrow)\rangle =
\sqrt{\frac{1}{6}}\left|\pm\frac{5}{2}\right\rangle-\sqrt{\frac{5}{6}}\left|
\mp\frac{3}{2}\right\rangle,\\
&&|\Gamma_8:(+\uparrow,+\downarrow)\rangle =
\sqrt{\frac{5}{6}}\left|\pm\frac{5}{2}\right\rangle+\sqrt{\frac{1}{6}}\left|
\mp\frac{3}{2}\right\rangle,\\
&&|\Gamma_8:(-\uparrow,-\downarrow)\rangle = \left|\pm\frac{1}{2}\right\rangle,
\end{eqnarray}
\begin{flushleft}
(ii) $j=7/2$
\end{flushleft}
\begin{eqnarray}
&&|\Gamma_6:(\uparrow,\downarrow)\rangle =
\pm\sqrt{\frac{5}{12}}\left|\mp\frac{7}{2}\right\rangle\pm\sqrt{\frac{7}{12}}\left|\pm\frac{1
}{2}\right\rangle,\\
&&|\Gamma_7:(\uparrow,\downarrow)\rangle =
\pm\sqrt{\frac{9}{12}}\left|\pm\frac{5}{2}\right\rangle\mp\sqrt{\frac{3}{12}}\left|\mp\frac{3
}{2}\right\rangle,\\
&&|\Gamma_8:(+\uparrow,+\downarrow)\rangle =
\pm\sqrt{\frac{3}{12}}\left|\pm\frac{5}{2}\right\rangle\pm\sqrt{\frac{9}{12}}\left|\mp\frac{3
}{2}\right\rangle,\\
&&|\Gamma_8:(-\uparrow,-\downarrow)\rangle =
\pm\sqrt{\frac{7}{12}}\left|\mp\frac{7}{2}\right\rangle\mp\sqrt{\frac{5}{12}}\left|\pm\frac{1
}{2}\right\rangle,
\end{eqnarray}
where irrelevant quantum numbers have been omitted.
To specify the components of irrep, we have used the symbols $\uparrow$ and $\downarrow$ for the time-reversal partner and the extra orbital labels $\pm$ for the $\Gamma_8$ irrep.
The relative phases of the basis are chosen so that the Kramers pair transforms like the spin 1/2 under the time-reversal operation $\theta$, i.e., $\theta|\uparrow\rangle=|\downarrow\rangle$, $\theta|\downarrow\rangle=-|\uparrow\rangle$.

Before we express the irreducible tensors, it is convenient to introduce two pseudospins $\sigma^\alpha$ and $\tau^i$ (each of them is defined as usual convention of Pauli matrices) which act on the Kramers and the non-Kramers pairs, respectively, without changing the other degrees of freedom \cite{uimin,shiina,fukushima,ohkawa}.
The six pseudospin operators are classified by the time-reversal operation.
Due to the definition of the pseudospin and the basis, it is easy to obtain their property under the time reversal:
\begin{eqnarray}
&&{\rm odd\;\;(magnetic)} \;:\; \sigma^x,\sigma^y,\sigma^z,\tau^y,\nonumber\\
&&{\rm even\;\;(electric)} \;:\; \tau^x,\tau^z.
\label{eqpa}
\end{eqnarray}
Notice that the pure imaginary operator $\tau^y$ has the same transformation property as magnetic moment.

Once we fix the basis sets, we can express the irreducible tensors \cite{butler,griffith} in a concise way by using the pseudospins.
We define $\sigma^0$ and $\tau^0$ as unit matrices in the spin and the orbital spaces, respectively, and the linear combinations of $\tau^x$ and $\tau^z$ as
\begin{eqnarray}
&& \eta^\pm = \frac{1}{2}(\pm\sqrt{3}\tau^x-\tau^z), \\
&& \zeta^\pm = -\frac{1}{2}(\pm\sqrt{3}\tau^z+\tau^x).
\end{eqnarray}
All tensors for possible combination of the basis sets are given as follows:
\begin{flushleft}
(i) $\Gamma_6\times\Gamma_6$
\end{flushleft}
\FL
\begin{equation}
\Gamma_{4m}\; :\; \left[\;\sigma^x,\sigma^y,\sigma^z\;\right],
\end{equation}
\begin{flushleft}
(ii) $\Gamma_7\times\Gamma_7$
\end{flushleft}
\FL
\begin{equation}
\Gamma_{4m} \; :\; \left[\;\sigma^x,\sigma^y,\sigma^z\;\right],
\end{equation}
\begin{flushleft}
(iii) $\Gamma_8\times\Gamma_8$
\end{flushleft}
\FL
\begin{eqnarray}
\Gamma_{2m} \; &:&\; \left[\;\tau^y\sigma^0 \;\right],\nonumber\\
\Gamma_{3e} \; &:&\; \left[\;\tau^z\sigma^0,\tau^x\sigma^0\;\right],\nonumber\\
\Gamma_{4m}^{(1)} \;&:&\; \left[\;\tau^0\sigma^x,\tau^0\sigma^y,\tau^0\sigma^z\;\right],\nonumber\\
\Gamma_{4m}^{(2)} \;&:&\; \left[\;\eta^+\sigma^x,\eta^-\sigma^y,\tau^z\sigma^z \;\right],\nonumber\\
\Gamma_{5m} \;&:&\; \left[\;\zeta^+\sigma^x,\zeta^-\sigma^y,\tau^x\tau^z \;\right],\nonumber\\
\Gamma_{5e} \;&:&\; \left[\;\tau^y\sigma^x,\tau^y\sigma^y,\tau^y\sigma^z \;\right],
\label{eq88}
\end{eqnarray}
\begin{flushleft}
(iv) $\Gamma_6\times\Gamma_7$
\end{flushleft}
\FL
\begin{eqnarray}
\Gamma_{2e} \;&:&\;
\left[\left(\begin{array}{c|c}
0 & \sigma^0\\ \hline \sigma^0 & 0
\end{array}\right)\right]
\nonumber\\
\Gamma_{5m} \;&:&\;
\left[
\left(\begin{array}{c|c}
0 & \sigma^x\\ \hline \sigma^x & 0
\end{array}\right),
\left(\begin{array}{c|c}
0 & \sigma^y\\ \hline \sigma^y & 0
\end{array}\right),
\left(\begin{array}{c|c}
0 & \sigma^z\\ \hline \sigma^z & 0
\end{array}\right)
\right],
\end{eqnarray}
\begin{flushleft}
(v) $\Gamma_6\times\Gamma_8$
\end{flushleft}
\FL
\begin{eqnarray}
\Gamma_{3e} \;&:&\;
\sqrt{\frac{3}{2}}\left[
\left(\begin{array}{c|cc}
0 & 0 & \sigma^0 \\ \hline 0 & 0 & 0 \\ \sigma^0 & 0 &  0
\end{array}\right),
\left(\begin{array}{c|cc}
0 & \sigma^0 & 0 \\ \hline \sigma^0 & 0 & 0 \\ 0 & 0 & 0
\end{array}\right)
\right]
\nonumber\\
\Gamma_{4m} \;&:&\;
\sqrt{\frac{3}{2}}\left[
\frac{1}{2}\left(\begin{array}{c|cc}
0 & -\sqrt{3}\sigma^x & \sigma^x \\ \hline
-\sqrt{3}\sigma^x & 0 & 0 \\ \sigma^x & 0 & 0
\end{array}\right),
\right.\nonumber\\&& \left.
\frac{1}{2}\left(\begin{array}{c|cc}
0 & \sqrt{3}\sigma^y & \sigma^y \\ \hline
\sqrt{3}\sigma^y & 0 & 0 \\ \sigma^y & 0 & 0
\end{array}\right),
-\left(\begin{array}{c|cc}
0 & 0 & \sigma^z \\ \hline 0 & 0 & 0 \\ \sigma^z & 0 & 0
\end{array}\right)
\right],\nonumber\\
\Gamma_{5m} \;&:&\;
\sqrt{\frac{3}{2}}\left[
\frac{1}{2}\left(\begin{array}{c|cc}
0 & \sigma^x & \sqrt{3}\sigma^x \\ \hline
\sigma^x & 0 & 0 \\ \sqrt{3}\sigma^x & 0 & 0
\end{array}\right),
\right.\nonumber\\&& \left.
\frac{1}{2}\left(\begin{array}{c|cc}
0 & \sigma^y & -\sqrt{3}\sigma^y \\ \hline
\sigma^y & 0 & 0 \\ -\sqrt{3}\sigma^y & 0 & 0
\end{array}\right),
-\left(\begin{array}{c|cc}
0 & \sigma^z & 0 \\ \hline \sigma^z & 0 & 0 \\ 0 & 0 & 0
\end{array}\right)
\right],
\nonumber\\&&
\end{eqnarray}
\begin{flushleft}
(vi) $\Gamma_7\times\Gamma_8$
\end{flushleft}
\FL
\begin{eqnarray}
\Gamma_{3e} \;&:&\;
\sqrt{\frac{3}{2}}\left[
\left(\begin{array}{c|cc}
0 & \sigma^0 & 0 \\ \hline \sigma^0 & 0 & 0 \\ 0 & 0 & 0
\end{array}\right),
-\left(\begin{array}{c|cc}
0 & 0 & \sigma^0 \\ \hline 0 & 0 & 0 \\ \sigma^0 & 0 & 0
\end{array}\right)
\right]
\nonumber\\
\Gamma_{4m} \;&:&\;
\sqrt{\frac{3}{2}}\left[
\frac{1}{2}\left(\begin{array}{c|cc}
0 & \sigma^x & \sqrt{3}\sigma^x \\ \hline
\sigma^x & 0 & 0 \\ \sqrt{3}\sigma^x & 0 & 0
\end{array}\right),
\right.\nonumber\\&&\left.
\frac{1}{2}\left(\begin{array}{c|cc}
0 & \sigma^y & -\sqrt{3}\sigma^y \\ \hline
\sigma^y & 0 & 0 \\ -\sqrt{3}\sigma^y & 0 & 0
\end{array}\right),
-\left(\begin{array}{c|cc}
0 & \sigma^z & 0 \\ \hline \sigma^z & 0 & 0 \\ 0 & 0 & 0
\end{array}\right)
\right],\nonumber\\
\Gamma_{5m} \;&:&\;
\sqrt{\frac{3}{2}}\left[
\frac{1}{2}\left(\begin{array}{c|cc}
0 & -\sqrt{3}\sigma^x & \sigma^x \\ \hline
-\sqrt{3}\sigma^x & 0 & 0 \\ \sigma^x & 0 & 0
\end{array}\right),
\right.\nonumber\\&&\left.
\frac{1}{2}\left(\begin{array}{c|cc}
0 & \sqrt{3}\sigma^y & \sigma^y \\ \hline
\sqrt{3}\sigma^y & 0 & 0 \\ \sigma^y & 0 & 0
\end{array}\right),
-\left(\begin{array}{c|cc}
0 & 0 & \sigma^z \\ \hline 0 & 0 & 0 \\ \sigma^z & 0 & 0
\end{array}\right)
\right],
\end{eqnarray}
where the subscript $m$ represents $magnetic$ and $e$ does $electric$.

The exchange interaction can be cast into the form eq.(\ref{eq2}) by using the explicit form of the irreducible tensors.
The number $n_{\Delta}^{(t)}$ of independent coupling constants for each localized-state tensors is evaluated by counting possible combination of scattering channels in the same irrep.
Precisely, there are 56 in total: $n_{2m}=n_{5e}=3$, $n_{3e}=8$, $n_{5m}=10$ and
$n_{4m}^{(1)}=n_{4m}^{(2)}=16$.

We note that $f^1$-$f^0$ charge fluctuation involves only the scattering channels of ($j=5/2$, $\Gamma_8$) symmetry in the exchange interaction because of the selection rule, eq.(\ref{eqsel}).
Thus, if one ignores $f^1$-$f^2$ charge fluctuation, the exchange interaction reduces to that of the SU(4) Coqblin-Schrieffer(CS) model.

\subsection{Explicit Form of Scaling Equations}
In deriving the scaling equations explicitly, we restrict ourselves for simplicity to the case where the $f^1$-$f^0$ charge fluctuation dominates over the $f^1$-$f^2$ one.
In this case, the $f^1$-$f^2$ fluctuation gives two different corrections to the SU(4) CS model: (i) modification of coupling constants breaking the SU(4) symmetry, and (ii) generation of additional scattering channels besides ($j=5/2$, $\Gamma_8$).
The latter correction gives very low characteristic energy as compared with the one given by the $f^1$-$f^0$ fluctuation and hardly affect  the renormalization of the SU(4) CS model.
On the other hand, the former correction changes the renormalization flow qualitatively and gives rise to multiple characteristic energies even though it is small.

We take into account the effect of $f^1$-$f^2$ fluctuations only for such processes that are absent in the SU(4) CS model.
Then there appear  the exchange interaction with the irreps ${\Delta}=2m,3e,5m,5e,4m$ by the decomposition of $\Gamma_8\otimes\Gamma_8$.
Of these the last one ${\Delta}=4m$ has the multiplicity 2 which we distinguish by using the matrix $g_{4m}^{(rt)}$ with $r,t =1,2$.
We deal with the following exchange interaction:
\begin{eqnarray}
H_{\rm ex}=&&\frac{1}{4}\left[
\sum_{\Delta\neq 4m}
g_{\Delta}(xX)_{\Delta}
+\sum_{r,t= 1}^2 g_{4m}^{(rt)}(x_rX_{t})_{4m}
\right],
\end{eqnarray}
where we define the summation of the components as
\begin{equation}
(x_rX_t)_\Delta=\sum_\delta
\hat{x}^{(r)}_{\Delta\delta}(\xi'\xi)X^{(t)}_{\Delta\delta}(\phi_g),
\end{equation}
with the scattering channel, $\xi'=\xi=(j=5/2,\Gamma_8)$ and the local ground-state configuration $\phi_g=(f^1,{}^2F_{5/2},\Gamma_8)$.
It is noted that the SU(4) CS model is reproduced by setting all coupling constants as equal to $g$ except for $g_{4m}^{(12)}$, $g_{4m}^{(21)}$ ($=0$).

We use the formulas (\ref{eqf1}), (\ref{eqf}) and (\ref{eqse}) together with the definition of tensors (\ref{eq88}) to obtain the set of scaling equations:
\widetext
\begin{eqnarray}
&&\frac{\partial}{\partial\ell}g_{2m}=-(g_{3e}^2+3g_{4m}^{(22)}g_{5m})+\frac{1}{2}g_{2m}[2g_{3e}^2+3(g_{4m}^{(12) 2}+g_{4m}^{(22) 2}+g_{5m}^2)],\\
&&\frac{\partial}{\partial\ell}g_{3e}=-\frac{1}{2}[2g_{2m}g_{3e}+3g_{5e}(g_{4m}^{(22)}+g_{5m})]+\frac{1}{4}g_{3e}[2(g_{2m}^2+g_{3e}^2)+3(g_{4m}^{(12) 2}+g_{4m}^{(22)2}+g_{5m}^2+2g_{5e}^2)],\\
&&\frac{\partial}{\partial\ell}g_{4m}^{(11)}=-\frac{1}{4}[4g_{4m}^{(11) 2}+g_{4m}^{(22)2}-2(g_{4m}^{(12) 2}+g_{4m}^{(21)2})+6g_{4m}^{(22)}g_{5m}+g_{5m}^2+4g_{5e}^2]\nonumber\\&&\mbox{\hspace{3cm}}
+[g_{4m}^{(11) 3}-g_{4m}^{(12)}g_{4m}^{(21)}g_{4m}^{(22)}+g_{4m}^{(11)}(g_{4m}^{(21)2}+g_{4m}^{(22) 2}+g_{5e}^2+g_{5m}^2)],\\
&&\frac{\partial}{\partial\ell}g_{4m}^{(12)}=-\frac{1}{2}[g_{4m}^{(21)}(g_{4m}^{(22)}-3g_{5m})-2g_{4m}^{(11)}g_{4m}^{(12)}]
\nonumber\\&&\mbox{\hspace{3cm}}
+\frac{1}{4}[g_{4m}^{(12)3}-4g_{4m}^{(11)}g_{4m}^{(21)}g_{4m}^{(22)}+g_{4m}^{(12)}\{4g_{4m}^{(21) 2}+g_{4m}^{(22)2}+2(g_{2m}^2+g_{3e}^2+g_{5e}^2)+5g_{5m}^2\}],\\
&&\frac{\partial}{\partial\ell}g_{4m}^{(21)}=-\frac{1}{2}[g_{4m}^{(12)}(g_{4m}^{(22)}-3g_{5m}
)-2g_{4m}^{(11)}g_{4m}^{(21)}]+[g_{4m}^{(21)3}-g_{4m}^{(11)}g_{4m}^{(12)}g_{4m}^{(22)}+g_{4m}^{(21)}(g_{4m}^{(11) 2}+g_{4m}^{(12)2}+g_{5e}^2+g_{5m}^2)],\\
&&\frac{\partial}{\partial\ell}g_{4m}^{(22)}=-\frac{1}{2}[g_{4m}^{(12)}g_{4m}^{(21)}+g_{4m}^{(11)}(g_{4m}^{(22)}+3g_{5m})+2(g_{3e}g_{5e}+g_{2m}g_{5m})]\nonumber\\&&\mbox{\hspace{3cm}}
+\frac{1}{4}[g_{4m}^{(22)}\{4g_{4m}^{(11) 2}+g_{4m}^{(12)2}+2(g_{2m}^2+g_{3e}^2+g_{5e}^2)+5g_{5m}^2\}
+g_{4m}^{(22) 3}-4g_{4m}^{(11)}g_{4m}^{(12)}g_{4m}^{(21)}],\\
&&\frac{\partial}{\partial\ell}g_{5m}=-\frac{1}{2}[3(g_{4m}^{(11)}g_{4m}^{(22)}-g_{4m}^{(12)}
g_{4m}^{(21)})+2(g_{4m}^{(22)}g_{2m}+g_{3e}g_{5e})+g_{4m}^{(11)}g_{5m}]\nonumber\\&&\mbox{
\hspace{3cm}}
+\frac{1}{4}[g_{5m}\{4(g_{4m}^{(11) 2}+g_{4m}^{(21) 2})+5(g_{4m}^{(12) 2}+g_{4m}^{(22) 2})
+2(g_{2m}^2+g_{3e}^2+g_{5e}^2)\}+g_{5m}^3],\\
&&\frac{\partial}{\partial\ell}g_{5e}=-[2g_{4m}^{(11)}g_{5e}+g_{3e}(g_{4m}^{(22)}+g_{5m})]+
\frac{1}{2}[2g_{5e}^3+g_{5e}\{2(g_{4m}^{(11) 2}+g_{4m}^{(21) 2}+g_{3e}^2)+g_{4m}^{(12)
2}+g_{4m}^{(22) 2}+g_{5m}^2\}].
\end{eqnarray}
\narrowtext\noindent
We note that these apparently complicated expressions follow straightforwardly from eq.(\ref{eqse}).
The correctness of the expression has been checked by taking various limiting cases.

\subsection{Nature of scaling}
Let us discuss implication of the scaling equations.
The set of equations has eight stable as well as saddle-point fixed points, which are summarized in Table \ref{tbl1}.
Because all fixed points have the relations $g_{4m}^{(22)*}=g_{5m}^*$ and
$g_{4m}^{(12)*}=g_{4m}^{(21)*}=0$, the exchange interaction at fixed points reads
\begin{eqnarray}
&&H_{ex}=\frac{1}{4}J\sum_\alpha\sigma^\alpha_c
\sigma^\alpha_f+\frac{1}{4}\left[K_\perp\left(\tau^x_c \tau^x_f+\tau^z_c \tau^z_f\right)
+K_z \tau^y_c \tau^y_f
\right]
\nonumber\\&&\mbox{\hspace{1cm}}
+\frac{1}{4}\sum_\alpha\left[I_\perp\left(\tau^x_c \tau^x_f+\tau^z_c
\tau^z_f\right)+I_z\tau^y_c \tau^y_f\right]
\sigma^\alpha_c \sigma^\alpha_f,
\label{eqex}
\end{eqnarray}
\newpage\vspace*{14.7cm}\noindent
where the coupling constants have been redefined as follows:
\begin{eqnarray}
&&g_{4m}^{(11)}=J,\;\;g_{3e}=K_\perp,\;\;g_{2m}=K_z,
\nonumber\\&&
g_{4m}^{(22)}=g_{5m}=I_\perp,\;\;g_{5e}=I_z.
\end{eqnarray}
The pseudospins for conduction electrons are rewritten explicitly as
\begin{eqnarray}
&&\tau^i_c\sigma^\alpha_c=\sum_{k'k}\sum_{m'm}^{\pm}\sum_{\sigma'\sigma}^{\uparrow,\downarrow
}c^\dagger_{k'm'\sigma'}\rho^i_{m'm}\rho^\alpha_{\sigma'\sigma}c_{km\sigma},
\nonumber\\&&\mbox{\hspace{3cm}}
\;\;\;(i,\alpha=0,x,y,z),
\end{eqnarray}
where $\rho^{\alpha}$ with $\alpha =x,y$ or $z$ denote the Pauli matrices and $\rho^0$ does the unit matrix.
Similar definition is also used for the pseudospins for localized states.
In the redefined expression of exchange interaction (\ref{eqex}), difference in time-reversal characters of the irreducible tensors appear as the exchange anisotropy of pseudospins.
This kind of $\sigma$-$\tau$ double tensor exchange model has been studied in the literature.
However unnecessary imposition on the parameters caused ambiguous conclusion about the fixed points \cite{pang}.

The two fixed points in a given row in Table \ref{tbl1} are essentially the same with each other since the one with the upper sign changes into another with the lower sign by a unitary transformation which changes simultaneously the signs of the transverse couplings, $K_\perp$ and $I_\perp$.
The groups with and without a prime are related to each other by the transformation $(I_\perp, I_z)\leftrightarrow -(I_\perp,I_z)$.
This hidden symmetry will be discussed later.

In the absence of the last term in eq.(\ref{eqex}), the $\sigma$ and $\tau$ spaces are decoupled.
Thus the fixed points (i), (ii) and (iii) correspond to: (i) the non-Fermi-liquid (NFL) fixed point in $\sigma$ space, and (ii) the Ising and (iii) the NFL fixed points in $\tau$ space, respectively.

The last term couples $\sigma$ and $\tau$ spaces.
In the absence of $K_\perp$, the term of $I_z$ leads to the fixed point (vi) where the NFL appears only in $\sigma$ space with the coupling constants $J_{\rm eff}=J+I_z\tau^y_c\tau^y_f$, while the term of $I_\perp$ to the fixed point (vii).
In the presence of $I_z$ or $I_\perp$ together with the transverse coupling $K_\perp$, the coupling constants flow to the fixed point (viii) with the SU(4) symmetry.
It is known that the Coqblin-Schrieffer model does not have the NFL fixed point.
Then, the finite magnitude of the fixed-point coupling is an artifact of the third-order scaling.
The correct fixed point should be at $J^*=K_\perp^*=K_z^*=\infty$ and gives the local Fermi liquid.
The stability of the saddle-point fixed points against each type of perturbation is summarized in Table \ref{tbl2}.

\begin{figure}
\begin{center}
\epsfxsize=8cm \epsfbox{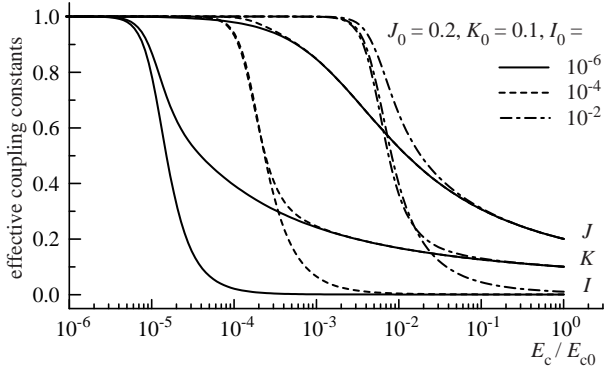}
\end{center}
\caption{The renormalization evolution of coupling constants for the initial coupling constants $J_0=0.2$ and $K_0=0.1$ with increasing $I_0=10^{-6}$, $10^{-4}$, and $10^{-2}$.
The three different characteristic energies corresponding to $J$, $K$ and $I$ rapidly merge together as $I_0$ increases.}
\label{fig3}
\end{figure}

Let us discuss renormalization evolution and characteristic energies for some simplified cases.
First, we consider the isotropic case: $K_\perp=K_z=K$ and $I_\perp=I_z=I$.
In this case, the scaling equations are reduced to those discussed in Ref. \cite{kuramoto2} as
\begin{eqnarray}
&&\frac{\partial}{\partial\ell}J=-(1-J)(J^2+3I^2),\\
&&\frac{\partial}{\partial\ell}K=-(1-K)(K^2+3I^2),\\
&&\frac{\partial}{\partial\ell}I=-2I(J+K)+I(J^2+K^2+2I^2).
\end{eqnarray}
In the absence of the coupling $I$, it is obvious that there exist two characteristic energies corresponding to each Kondo effects: $T_{\rm K}^J/E_c=J\exp(-1/J)$ and $T_{\rm K}^K/E_c=K\exp(-1/K)$.
On the other hand, the case of SU(4) symmetry with $I=J=K$ gives single characteristic energy, $T_{\rm K}^{\rm CS}/E_c=I^{1/4}\exp(-1/4I)$.
It is noted that $T_{\rm K}^{\rm CS}$ is larger than $T_{\rm K}^{J,K}$ since the number of the screening channels is larger.
The renormalization evolution of the coupling constants is shown in Fig.\ref{fig3} for the initial coupling constants $J_0=0.2$ and $K_0=0.1$ with  $I_0=10^{-6}$, $10^{-4}$ and $10^{-2}$.
The three different characteristic energies corresponding to $J$, $K$ and $I$ merge together as $I_0$ increases.
The average of the characteristic energies also increases as $I_0$ increases.

\begin{figure}
\begin{center}
\epsfxsize=8cm \epsfbox{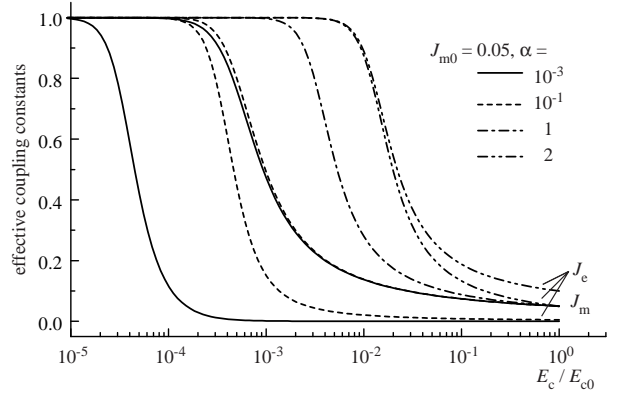}
\end{center}
\caption{The renormalization evolution of coupling constants for the initial magnetic coupling constant $J_{m0}=0.05$ with increasing the ratio $\alpha=J_{e0}/J_{m0}$, $J_e$ being the electric coupling constant.
The two different characteristic energies corresponding to $J_m$ and $J_e$  merge together as $\alpha$ approaches unity.}
\label{fig4}
\end{figure}

Next, we focus on the difference in time-reversal property and put $J_m=J=K_z=I_\perp$ and $J_e=K_\perp=I_z$.
The scaling equations are simplified in this case as
\begin{eqnarray}
&&\frac{\partial}{\partial\ell}J_m=-(1-J_m)(3J_m^2+J_e^2),\\
&&\frac{\partial}{\partial\ell}J_e=-2J_e[2J_m-(J_m^2+J_e^2)].
\end{eqnarray}
In the absence of the electric coupling $J_e$, the characteristic energy corresponding to the magnetic Kondo effect is given by $T_{\rm K}^m/E_c=J_m^{1/3}\exp(-1/3J_m)$.
$T_{\rm K}^m$ is also smaller than $T_{\rm K}^{\rm CS}$ since only the magnetic channels contribute to the Kondo effect.
The renormalization evolution is shown in Fig.\ref{fig4} for the initial coupling constants $J_{m0}=0.05$ with various values of the ratio $\alpha=J_{e0}/J_{m0}$.
The two different characteristic energies corresponding to $J_m$ and $J_e$ merge together as $\alpha$ approaches unity.

Before closing this section, we discuss the hidden symmetry $I\leftrightarrow-I$.
Let us consider the following particle-hole transformation for the conduction electrons:
\begin{equation}
c_{km\sigma}\rightarrow\sum_{i\alpha}T_{mi}S_{\sigma\alpha}c^\dagger_{-ki\alpha},
\end{equation}
where the matrices $S$ and $T$ are unitary.
The transformation does not change the kinetic-energy part of the Hamiltonian unless the conduction band is asymmetric with respect to the Fermi level.
On the other hand, the spin operators $\sigma^i_c$ and $\tau^\alpha_c$ are transformed respectively as
\begin{eqnarray}
&&\tau^i_c\rightarrow -\sum_{k'k}\sum_{\sigma
m'm}c^\dagger_{k'm'\sigma}\bar{\rho}^i_{m'm}c_{km\sigma},\\
&&\sigma^\alpha_c\rightarrow -\sum_{k'k}\sum_{m\sigma'\sigma}c^\dagger_{k'm\sigma'}\bar{\rho}
^\alpha_{\sigma'\sigma}c_{km\sigma},\\
&&\tau^i_c\sigma^\alpha_c\rightarrow
-\sum_{k'k}\sum_{m'm\sigma'\sigma}c^\dagger_{k'm'\sigma'}\bar{\rho}^i_{m'm}\bar{\rho}^\alpha_
{\sigma'\sigma}c_{km\sigma},
\end{eqnarray}
where the minus signs come from the anticommutation of conduction-electron operators.
The transformed spin operators with bars are given by
\begin{equation}
\bar{\rho}^i=[T^\dagger\rho^iT]^t,\;\;\;
\bar{\rho}^\alpha=[S^\dagger\rho^\alpha S]^t.
\end{equation}
Let us assume the presence of unitary transformations $S$ and $T$ which make the transformed spin operators proportional to the original one.
Namely, we have $\bar{\rho}^i=A\rho^i$ and $\bar{\rho}^\alpha=B\rho^\alpha$ for all components where $A$ and $B$ are independent of the component indices.
Then the transformed exchange interaction is equivalent to the original one with the coupling constants $\bar{J}=-AJ$, $\bar{K}=-BK$ and $\bar{I}=-ABI$.
There indeed exist such transformations in the case of the SU(2) symmetry; the matrices are given by $S=T=\rho^y$ which give $A=B=-1$.
However, in the case of SU($N$) symmetry with $N>2$,  such transformations cannot be found since the rank of the SU($N$) symmetry is higher than 2.
We note that the hidden symmetry for SU(2) plays a special role to reject the NFL fixed point which is present in the large limits of both $N$ and $M$ in the SU($N$)$\times$SU($M$) exchange model discussed in Ref. \cite{kuramoto2}.

%
%
\section{Summary}
In this paper we have proposed a systematic procedure to derive the exchange interaction and the scaling equations up to third order with maximum use of the point-group symmetry.
Various moments are described by the irreducible tensors with proper time-reversal property.
Then the exchange interaction is obtained in the scalar-product form of the irreducible tensors.
The third-order scaling equations for symmetry adapted coupling constants are written down generally in terms of structure constants of the relevalnt Lie algebra.

The procedure is applied to the case of Ce$_x$La$_{1-x}$B$_6$ where the CEF ground quartet $\Gamma_8$ shows rich phenomena resulting from the  entanglement of electric and magnetic tensors.
We derive the exchange interaction integrating out the charge fluctuations  to $f^0$ or $f^2$ configurations.
We discuss the nature of scaling for the case where the $f^1$-$f^0$ fluctuation dominates over the $f^1$-$f^2$ one.
In this case, main contribution to the Kondo screening comes from the partial waves with $j=5/2$, $\Gamma_8$ symmetry of conduction electrons.

As a result of scaling, the effective exchange interaction is described by the $\sigma$-$\tau$ double exchange model with the exchange anisotropy in $\tau$ space.
The anisotropy comes from the fact that one of three components of the orbital pseudospin $\tau^y$ has different time-reversal property from others.
The effective exchange model has non-Fermi-liquid fixed points in the absence of the $\sigma$-$\tau$ coupled term.
In the presence of the coupled term the system flows to the local Fermi-liquid fixed point.
The different characteristic energies corresponding to the exchange couplings for either $\sigma$ or and $\tau$ rapidly merge together with increasing magnitude of the coupling.
If one considers only the time-reversal difference, the system has also two different characteristic energies corresponding to the electric and the magnetic exchange interactions.
We identify the special symmetry of SU(2), which plays an important role to reject the non-Fermi-liquid fixed point resulting from the argument of SU($N$)$\times$SU($M$) exchange model in the large limit of $N$ and $M$ \cite{kuramoto2}.

Concerning the physical quantities in the intermediate regime toward the fixed point, magnetic and electric quantities can have different energy scales.
Therefore, conventional argument with the single Kondo temperature has to be revised in the presence of the orbital degeneracy.
This should affect previous analysis of the resistivity and magnetic susceptibility.
For this purpose,  quantitative argument is required beyond the perturbative scaling analysis.
In the forthcoming paper we use the numerical renormalization group \cite{kusunose} to derive physical quantities at finite temperature.

In the presence of intersite interactions, the orbital and spin degeneracies can be split by a molecular field associated with the long-range order.
It is known that Ce$_x$La$_{1-x}$B$_6$ has at least four different phases.
Of these the phase II is characterized by the orbital (or quadrupole) order without magnetic order.
The magnetic order present in the phase III seems to accompany the orbital order.
If the molecular field corresponds to the internal magnetic field, not only spin but orbital degeneracy are lifted because Zeeman splitting depends on each orbital.
Provided that the intersite spin exchange interaction is stronger than the orbital one, we can think of the situation where the spin Kondo effect gives way to the long-range order but the orbital Kondo effect does not.
In such a situation the magnetically ordered phase should keep the nearly isotropic magnetic property as in the paramagnetic phase.
We expect that a detailed quantitative study with the scenario described above will provide us with understanding the nature of the curious phase IV in Ce$_x$La$_{1-x}$B$_6$.

%
%
\acknowledgments
We would thank Professor O. Sakai for useful conversation on the symmetry classification.
One of the authors (H. K.) is also grateful to Professor P. Thalmeier and Dr. T.A. Costi for valuable comments.
This work was supported by Grand-in-Aid for encouragement of Young Scientists from the Ministry of Education, Science and Culture of Japan and also by CREST from the Japan Science and Technology Corporation.

\appendix
%
%
\section{Explicit Derivation of Exchange Interaction}
Here we derive the explicit expression of the coupling constants by applying the second-order perturbation to the Anderson model (\ref{eqand}).
The exchange interaction via the $n+1$ configuration is given by
\begin{eqnarray}
&&H_{\rm ex}^+=-\sum_{\xi'\xi\phi_n'\phi_n}\sum_{\phi_{n+1}}I_+\sum_{\gamma'\gamma
k'\lambda'k\lambda}c^\dagger_{k'\xi'\lambda'}c_{k\xi\lambda}|\phi_n'\gamma'\rangle\langle\phi_n\gamma|
\nonumber\\&&\mbox{\hspace{0.5cm}}\times
\sum_{\gamma_+}
\left(\begin{array}{ccc}
\Gamma_+^* & \Lambda' & \Gamma' \\ \gamma_+^* & \lambda' & \gamma'
\end{array}\right)^*_{q'}
\left(\begin{array}{ccc}
\Gamma_+^* & \Lambda & \Gamma \\ \gamma_+^* & \lambda & \gamma
\end{array}\right)_q,
\label{eqa1}
\end{eqnarray}
with
\begin{eqnarray}
&&I_+=\frac{V_{\xi'}V_\xi^*}{E_f(\phi_{n+1})-E_f(\phi_n)}
\nonumber\\&&\mbox{\hspace{1cm}}\times
\langle\phi_n'||f_{\xi'}||\phi_{n+1}\rangle\langle\phi_{n+1}||f^\dagger_\xi||\phi_n\rangle.
\end{eqnarray}
The reduced matrix elements of the creation operator can be calculated by using e.g. the Racah factorization lemma in addition to the coefficient of fractional parentage \cite{butler,nielson}.

The summation of the products of the $3jm$ symbols in eq.(\ref{eqa1}) can be converted to the summation of another combinations of $3jm$ symbols which relate to the product of the irreducible tensors \cite{hirst}.
For this purpose the following identity is useful:
\begin{eqnarray}
&&\sum_{\gamma_+}
\left(\begin{array}{ccc}
\Gamma_+^* & \Lambda' & \Gamma' \\ \gamma_+^* & \lambda' & \gamma'
\end{array}\right)^*_{q'}
\left(\begin{array}{ccc}
\Gamma_+^* & \Lambda & \Gamma \\ \gamma_+^* & \lambda & \gamma
\end{array}\right)_q=
\nonumber\\&&
\sum_{\Delta\delta rt}
\left\{\begin{array}{ccc}
\Delta^* & \Lambda'^* & \Lambda \\ \Gamma_+ & \Gamma & \Gamma'
\end{array}\right\}_{tq'qr}|\Delta|\{\Gamma_+^*\Lambda\Gamma q\}\{\Gamma^*\Delta^*\Gamma' t\}\{\Gamma^*\}
\times\nonumber\\&&
\left(\begin{array}{c} \Delta \\ \delta \end{array}\right)
\left(\begin{array}{c} \Lambda' \\ \lambda' \end{array}\right)
\left(\begin{array}{ccc}
\Lambda'^* & \Delta^* & \Lambda \\ \lambda'^* & \delta^* & \lambda
\end{array}\right)_r
\left(\begin{array}{c} \Gamma' \\ \gamma' \end{array}\right)
\left(\begin{array}{ccc}
\Gamma'^* & \Delta & \Gamma \\ \gamma'^* & \delta & \gamma
\end{array}\right)_t,
\nonumber\\
\end{eqnarray}
where the $6j$ symbol in the point-group irrep has appeared together with the $2j$ and $3j$ phases \cite{butler}.
By using the definitions of the irreducible tensors, eq.(\ref{eqten}), we obtain the coupling constants for the $n+1$ process as
\begin{eqnarray}
&&g^{(rt)}_{\Delta+}=\sum_{\phi_{n+1}}I_+
\left\{\begin{array}{ccc}
\Delta^* & \Lambda'^* & \Lambda \\ \Gamma_+ & \Gamma & \Gamma'
\end{array}\right\}_{tq'qr}|\Delta|(-1)^P,
\end{eqnarray}
with the phase
\begin{equation}
(-1)^P=-\{\Gamma_+^*\Lambda\Gamma q\}\{\Gamma^*\Delta^*\Gamma' t\}\{\Gamma^*\}.
\end{equation}

Similar argument for the $n-1$ configuration provides the coupling constants for the $n-1$ process:
\begin{eqnarray}
&&g^{(rt)}_{\Delta-}=\sum_{\phi_{n-1}}I_-
\left\{\begin{array}{ccc}
\Delta^* & \Lambda & \Lambda'^* \\ \Gamma_- & \Gamma & \Gamma'
\end{array}\right\}_{tqq'r}|\Delta|(-1)^{P'},
\end{eqnarray}
with
\begin{eqnarray}
&&I_-=\frac{V_{\xi'}V_\xi^*}{E_f(\phi_{n-1})-E_f(\phi_n)}
\nonumber\\&&\mbox{\hspace{1cm}}\times
\langle\phi_n'||f^\dagger_\xi||\phi_{n-1}\rangle\langle\phi_{n-1}||f_{\xi'}||\phi_n\rangle,
\\&&
(-1)^{P'}=
\{\Delta^*\}\{\Lambda^*\}\{\Gamma^*\}
\nonumber\\&&\mbox{\hspace{1cm}}\times
\{\Gamma_-\Gamma'^*\Lambda q\}\{\Lambda^*\Delta\Lambda' r\}\{\Gamma'^*\Delta\Gamma t\}.
\end{eqnarray}
Thus, the full form of the exchange interaction is given by
\begin{eqnarray}
&&H_{\rm ex} =
\sum_{\xi'\xi\phi_n'\phi_n}\sum_{\Delta rt}
g^{(rt)}_{\Delta}(\xi'\xi;\phi_n'\phi_n)
\nonumber\\&&\mbox{\hspace{2cm}}\times
\sum_\delta
\left(\begin{array}{c} \Delta \\ \delta \end{array}\right)
x^{(r)}_{\Delta^*\delta^*}(\xi'\xi)X^{(t)}_{\Delta\delta}(\phi_n'\phi_n),
\\&&\mbox{\hspace{1cm}}
g^{(rt)}_{\Delta}(\xi'\xi;\phi_n'\phi_n)=g^{(rt)}_{\Delta+}+g^{(rt)}_{\Delta-}.
\end{eqnarray}

%
%

\newpage
%
%

%
%
\begin{table}
\caption{The stable and the saddle-point fixed points for scaling equations.
The coupling constants in the redefined notation are written in the second line.}
\begin{tabular}{lcccccc}
&$g_{4m}^{(11)*}$&$g_{3e}^*$&$g_{2m}^*$&$g_{4m}^{(22)*}=g_{5m}^*$&$g_{5e}^*$&$g_{4m}^{(12)*}=
g_{4m}^{(21)*}$\\
& $J$ & $K_\perp$ & $K_z$ & $I_\perp$ & $I_z$ & \\ \tableline
(i)       &    1 &      0 &           0 &      0 &    0 &    0\\
(ii)      &    0 &      0 &    $K_{z0}$ &      0 &    0 &    0\\
(iii)     &    0 & $\pm1$ &           1 &      0 &    0 &    0\\
(iv)      &    1 &      0 &    $K_{z0}$ &      0 &    0 &    0\\
(v)       &    1 & $\pm1$ &           1 &      0 &    0 &    0\\
(vi)      &    1 &      0 &    $K_{z0}$ &      0 & $+1$ &    0\\
(vi')     &    1 &      0 &    $K_{z0}$ &      0 & $-1$ &    0\\
(vii)     &    1 &      0 &           1 & $\pm1$ &    0 &    0\\
(viii)    &    1 & $\pm1$ &           1 & $\pm1$ & $+1$ &    0\\
(viii')   &    1 & $\pm1$ &           1 & $\mp1$ & $-1$ &    0
\end{tabular}
\label{tbl1}
\end{table}

\begin{table}
\caption{The destinations of the saddle-point fixed points (i)--(vi') in the first column against each type of perturbation.}
\begin{tabular}{lccccccc}
& $J>0$ & $J<0$ & $K_\perp$ & $K_z$ & $I_\perp$ & $I_z>0$ & $I_z<0$\\ \tableline
(i)   &   -  &   -   &  (v)    & (iv) & (vii)   & (vi)   & (vi')   \\
(ii)  & (iv) & (ii)  & (iii)   &    - & (vii)   & (vi)   & (vi')   \\
(iii) & (v)  & (iii) &   -     &    - & (viii)  & (viii) & (viii') \\
(iv)  & -    &    -  & (v)     & -    & (vii)   & (vi)   & (vi')   \\
(v)   & -    &    -  & -       & -    & (viii)  & (viii) & (viii') \\
(vi)  & -    &    -  & (viii)  & -    & (viii)  & -      & -       \\
(vi') & -    &    -  & (viii') & -    & (viii') & -      & -       \\
(vii) & -    &    -  & (viii)  & -    & -       & (viii) & (viii')
\end{tabular}
\label{tbl2}
\end{table}

\end{document}